# Photon transport enhanced by transverse Anderson localization in disordered superlattices

P. Hsieh[1,*], C. Chung[2], J. F. McMillan[1], M. Tsai[3], M. Lu[4], N. C. Panoiu[5,6,*], and C. W. Wong[1,7,*]

[1] Optical Nanostructures Laboratory, Center for Integrated Science and Engineering, Solid-State Science and Engineering, Mechanical Engineering, Columbia University, New York, NY 10027, USA

[2] Center for Micro/Nano Science and Technology and Advanced Optoelectronic Technology Center, National Cheng Kung University, Tainan 701, Taiwan

[3] Center for Measurement Standards, Industrial Technology Research Institute, Hsinchu 300, Taiwan

[4] Center for Functional Nanomaterials, Brookhaven National Laboratory, Upton, NY 11973, USA

[5] Department of Electronic and Electrical Engineering, University College London, Torrington Place, London WC1E 7JE, UK

[6] Thomas Young Centre, London Centre for Nanotechnology, University College London, 17-19 Gordon Street, London, WC1H 0AH, UK

[7] Mesoscopic Optics and Quantum Electronics Laboratory, University of California, Los Angeles, CA 90095, USA

* Correspondence and requests for materials should be addressed to P.H., N.C.P. and C.W.W. (email: ph2285@columbia.edu, npanoiu@ee.ucl.ac.uk, and cheewei.wong@ucla.edu)

**One of the daunting challenges in optical physics is to accurately control the flow of light at the subwavelength scale, the main impediment being limitations from diffraction. Negative or zero index of refraction, transformational cloaking, metamaterials, and slow-light are a few such unique functionalities, including recent photon Anderson localization in disordered media, a ubiquitous wave phenomena. Here we report the photon transport and collimation enhanced by transverse Anderson localization in chip-scale dispersion engineered anisotropic media. We demonstrate a new type of photonic crystal superlattice structure in which diffraction is nearly completely arrested by cascaded resonant tunneling through transverse guided resonances. By modifying the geometry of more than 4,000 scatterers in the superlattices we add structural disorder controllably and uncover the mechanism of disorder-induced transverse localization in the solid-state. Arrested spatial divergence is captured in the power-law scaling, along with exponential asymmetric mode**

**profiles and enhanced collimation bandwidth for increasing disorder. With increasing disorder, we observe the crossover from cascaded guided resonances into the transverse localization regime, beyond both the ballistic and diffusive transport of photons.**

In regular isotropic optical media the characteristics of dispersion relations, which among others define the properties of diffraction, are determined by the intrinsic structure of the medium so that there is little room to engineer the optical wave diffraction. By contrast, structuring the optical medium at the subwavelength scale can lead to dramatic changes of the characteristics of dispersion and wave diffraction. One such salient example is that of photonic crystals [1-9], whose wave dispersion and diffraction are engineered so as to achieve specific functionalities. Drawing analogy to the transport of electrons in crystal solids, photonic crystals are recognized to provide insights of localization [10] in disordered and periodic scattering lattices. In particular, light localization in disordered media including that of transverse localization in optically-induced lattices [11] has been intensely investigated over the past years [12-21]. For monochromatic electromagnetic propagation in an inhomogeneous and nondissipative dielectric medium, wave transport with time-harmonic electric field amplitude $\vec{E}$ can be described by a Schrödinger-like equation:

$$-\nabla^2 \vec{E} + \vec{\nabla}(\vec{\nabla} \cdot \vec{E}) - \frac{\omega^2}{c^2}\varepsilon_{fluct}(z)\vec{E} = \varepsilon_o \frac{\omega^2}{c^2}\vec{E} \qquad (1)$$

where the dielectric scattering potential fluctuations $\varepsilon_{fluct}$ are distinct from the background (periodic) potential $\varepsilon_o$, such as in tight-binding models for disordered electronic transport. For electrons in the weak disorder limit (root-mean-square potential fluctuations $V_{rms}$ less than $\hbar^2/2m^*a_c^2$ where $m^*$ is the effective mass and $a_c$ the fluctuations correlation length), a Mott transition can occur [22]; in the strong disorder limit ($V_{rms}$ greater than $\hbar^2/2m^*a_c^2$), an Anderson transition [10] can occur for near-universal localization in real materials. Such localization transitions for photons are also possible with strong disorder, examined prior in the longitudinal on-axis propagating direction [23]. Furthering the electronic-photon analogy, the scaling theory of localization (zero conductance Σ for long lengthscales in 1D and 2D, and the mobility edge in 3D) and a modified Ioffe-Regel criteria ($kl^* \approx 1$ , where $k$ is the Bloch wavevector and $l^*$ the scattering mean free path) are also relevant in electromagnetic transport.

However, unlike electron transport where localized bound states are in deep potential wells, photon localization is at an intermediate frequency band (between low-frequency Rayleigh extended states and high-frequency geometric optics propagation) and at an energy *higher* than the highest potential wells [12-13]. As illustrated in Equation (1), the electromagnetic field is also vectorial and has an additional polarization density term $\vec{\nabla} \cdot \vec{E}$ that has no electronic analogue. Furthermore, working with photons, photonic lattices offer an unequivocal scaling test of localization in a static disorder potential, unhindered by many-body electron-electron and electron-phonon scattering, as one of the most accessible approaches to examine localization. Examples include the first observations of photon transverse localization in bulk photorefractive crystals [10] which, with the $\sim 5 \times 10^{-4}$ index contrast in the paraxial limit, can be described by $i\frac{\partial A}{\partial z} + \frac{1}{2k}\left(\frac{\partial^2 A}{\partial x^2} + \frac{\partial^2 A}{\partial y^2}\right) + \frac{k_T}{n_o}\Delta n(x,y)A = 0$, where $A(\vec{r})$ is the slowly-varying envelope of the time-harmonic field and $k_T$ the transverse wavevector. With strong index contrast ($\sim 2$) on-chip, however, intensive direct numerical approaches on Maxwell's equations have to be performed, with recent computational models of the pseudogap spectral function and photon density of states $\rho(\omega)$ in the band edge vicinity, through for example Bloch-mode expansion approaches [24]. Coherent backscattering in localization has been examined numerically and experimentally [25], supporting the possibility of the scaling theory of localization on-chip. Guided resonances in superlattices have also been modeled numerically and observed experimentally [2]. In these high-index disordered superlattices, we postulate that the resulting transverse guided resonances, with disorder-induced inhomogeneous spectral broadening, can potentially provide improved collimation bandwidth while experiencing, within this frequency range, transverse localization.

Figure 1 shows the nanofabricated chip-scale anisotropic superlattices examined in our study, consisting of alternating layers of photonic crystal sections of thickness $d_1$, made of circular holes arranged in a two-dimensional hexagonal lattice with lattice constant, $a = 500$ nm, and homogeneous sections of medium with thickness $d_2$ for a superperiod, $\Lambda = d_1 + d_2$. To introduce structural disorder, three other structures are also nanofabricated with heptagonal-hole superlattices (HHS; $\approx 2\%$ disorder), square-hole superlattices (SHS; $\approx 6\%$ disorder), and triangular-hole superlattices (THS; $\approx 13\%$ disorder). In each of these superlattices, disorder is introduced by randomly rotating each scatterer with a stochastically-uniform distribution of the rotation angle. All devices are fabricated in silicon-on-insulator (see Methods), with 20 superperiods, and the incoming transverse-magnetic-like (TM-like) polarized light is coupled into the superlattices via a single-mode waveguide of width, $w = 450$ nm.

The thickness of the homogeneous section satisfies the relation $d_2/d_1 = 0.18$, with the superlattice band structures, computed along the $\Gamma$-$X_2$ direction of the superlattice, shown in Figure 1f and Supplementary Information I. Significantly, as the superlattice band structure suggests, our photonic structure possesses nearly flat bands (highlighted in red) at the normalized frequencies of 0.314 and 0.327 (in dimensionless units of $\omega a/2\pi c$; centered around 0.322) at $k_x = 0$, corresponding to the high-symmetry $\Gamma$ point. These flat bands represent leaky guided resonances (located outside the light cone) [3], which propagate transversely in the 1D homogeneous dielectric region that forms a 1D photonic crystal waveguide (in dashed white lines in Figures 1b to 1d) separating the photonic crystal sections of the superlattices. The underlying mechanism that leads to enhanced collimation in these superlattices is as follows: mutual coupling of the two leaky guided resonances excited at the input and output interfaces of a homogeneous section gives rise to the mode splitting seen in the highlighted-red bands of Figure 1f. The $|E|^2$-field profiles of these resonances are shown in the insets of Figure 1f. Bloch modes of the photonic crystal couple to these guided resonances and are resonantly amplified when tunneling from one photonic crystal section to the next. This mechanism of resonant wave tunneling via excitation of guided resonances enhances the diffraction-free beam collimation since the evanescent part of the optical field is propagated through the superlattice as well. This beam collimation mechanism based on resonant tunneling – from guided resonances to guided resonances – is markedly different from that investigated in earlier studies [6-8], in which case the beam divergence is reduced by designing flat spatial dispersion surfaces or by alternating metamaterials layers of normal and anomalous dispersion (see Supplementary Information I to III for detailed design of the superlattices).

To quantify the degree of beam collimation, we illustrate in Figure 2 the computed effective beam width, $\omega_{effc} = P^{-1}$, defined as the inverse participation ratio, $P(z) \equiv [\int I(x,z)^2 dx]/[\int I(x,z) dx]^2$ [10], where $I(x,z)$ is the field intensity. We employed in these calculations 3D finite-difference time-domain (FDTD) simulations (see Methods and Supplementary Information IV), performed across the 1500 nm to 1600 nm spectral domain with 5 nm resolution. The blue regions in Figure 2 illustrate the regions of tightest collimation; for our designed CHS, the collimation band is centered at 1550 nm. With increasing disorder, the HHS, SHS, and THS structures show significantly larger bandwidths for collimation as compared to the CHS. This is attributed to the inhomogeneous spectral broadening of the guided resonances induced by disorder. The frequency of the guided resonances at the $\Gamma$ point is shifted by a random amount due to the coupling of the optical mode with the adjacent, randomly perturbed photonic crystal sections of the superlattices [24], an effect that is also accompanied by increased radiation losses. Since the frequency dispersion of these tunneling channels increases with disorder, the enhanced collimation bandwidth increases with disorder level as well. We note that in the instance of the rotated SHS the spectral region of strong collimation is slightly blue-shifted compared to the CHS due to the fact that, even if the hole area is kept the same, the frequency dispersion of the guided resonances depends weakly on the hole shape.

Encouraged by these theoretical predictions, we examined the far-field infrared scattering for 900 wavelengths (1530 nm to 1620 nm with 100 pm spectral resolution), for each of the superlattices. Figure 3a highlights the key features, with additional supporting examples shown in the Supplementary Information V. For the CHS, the most effective collimation is observed at 1550 nm ($\lambda_{ec}$), with the beam width at the interfaces, $\omega_{FWHM,i}$, fluctuating by less than $\pm\,7\%$. This wavelength is closest to that of the guided resonances, allowing for more effective coupling, with larger tunneling transmission and amplification of the evanescent part of the field. This is supported by the spectral analysis of the spatial full-width half-maximum (FWHM), $\omega_{FWHM}$, with the smallest beam width observed at $\lambda_{ec}$ and matching well with the numerical simulations data, where the strongest collimation occurs around $\lambda_{ec} - 4$ to $\lambda_{ec} + 4$ nm.

Figure 3a also illustrates the electromagnetic propagation for the disordered HHS, SHS, and THS cases at the corresponding $\lambda_{ec}$ wavelengths, compiled from 2,700 scattering images. Collimation is observed even in these disordered superlattices. The most effective $\lambda_{ec}$ wavelengths are determined to be $\approx$ 1550 nm (HHS), 1555 nm (SHS), and 1580 nm (THS) respectively. At other wavelengths, the beam diverges significantly from its input excitation width in the disordered superlattices. Concurrently the larger disorder superlattices such as the triangular and square realizations show shorter transmission lengths due to the increased disorder scattering losses from the perturbed Bloch modes. To observe finer features in the *z*-direction, we next perform near-field scanning optical microscopy (NSOM) at the $\lambda_{ec}$ wavelengths to probe the local field intensity oscillations in each superlattice (see Supplementary Information VI). Mapping the near-field intensity with the superimposed photonic crystal topography, the periodic enhancement of the wave scattering is determined to be centered at the location of the transverse waveguides. These near-field measurements (calibrated with a periodic topography grid) also show the *z*-thin *x*-long scattering slices corresponding to the thin homogeneous transverse waveguides. With increasing disorder, the near-field intensities at the interfaces become increasingly apparent compared to the background stray light (see Figure S13) due to the increased scattering into the radiation continuum and the more efficient excitation of the guided modes.

To compare against the guided resonances approach, we next designed photonic crystals with sizes of a few hundred micrometers [6-8] but without the superlattices and with flat equifrequency dispersion curves. We nanofabricated and examined, under the same conditions, collimation in these lattices, as detailed in Supplementary Information VII. Figure 3b illustrates the observed beam propagation at $\lambda_{ec}$ in the presence of disorder, in the collimation regime. The field profiles in Fig. 3b clearly demonstrate that in this case beam collimation is of a markedly different nature, as it almost completely vanishes in the presence of disorder. The averaged collimating beam width increases from ≈ 2.2 μm to ≈ 2.5 μm (heptagon-hole), ≈ 6.8 μm (square-hole), and ≈ 13.9 μm (triangular-hole), without the guided resonance contributions. The fluctuation of the beam width increases from ± 5%, to ± 6% (heptagon-hole), ± 9% (square-hole), and ± 11% (triangular-hole).

Figures 4a illustrates the optical wave transport in the superlattices at different wavelengths, for different disorder. The physical nature of the electromagnetic propagation is revealed by the slope of the function $\omega_{FWHM}(z)$ when represented on a log-log scale. As illustrated in the log-log plots of Figures 4b to 4e, the asymptotic dependence of the experimental effective $\omega_{FWHM}$ is of the form $\omega_{FWHM}(z) \propto z^\nu$, where the slope ν is a power exponent determined by linear fitting. For the CHS in Figure 4b, we observe ν values up to 0.24 at the longer wavelengths, but with a near-zero slope ν of ≈ 0.05 between $\lambda_{ec}$ – 4 to $\lambda_{ec}$ + 4 nm. This corresponds to ≈ 8 nm collimation bandwidth and is due solely to the beam interaction with the guided resonances. In the presence of ≈ 2% and ≈ 6% structural disorder (HHS and SHS, respectively), however, the measured log-log plots of $\omega_{FWHM}(z)$ show markedly different spectral dependence. The slope ν decreases significantly in the HHS between $\lambda_{ec}$ – 4 to $\lambda_{ec}$ + 17 nm, and in the SHS between $\lambda_{ec}$ – 4 to $\lambda_{ec}$ + 17 nm. This is illustrated in Figures 4c and 4d respectively. In both superlattices a near-zero ν value of ≈ 0.05 is now achieved within a 21 nm collimation bandwidth, sizably larger than in the CHS.

The observed increased collimation arises from the disorder-induced inhomogeneous spectral broadening of guided resonances. To further support this, we next examined the THS structure, with larger ($\approx$ 13%) structural disorder. The analyzed experimental collimation bandwidth is even larger, namely $\approx$ 31 nm ($\lambda_{ec}$ – 7 to $\lambda_{ec}$ + 24 nm) as illustrated in Figure 4e. These observed near-zero ν bandwidths are also larger than (and outside) the bandwidth of the computed regular CHS without disorder, with a $\approx$ 3.9× increase in collimation bandwidth achieved experimentally in the presence of disorder as compared to the periodic disorder-free CHS. We also note that, to characterize the effects of disorder, an ensemble average is needed over different realizations of disorder; in our superlattices the ensemble average is self-consistently performed as the beam propagates over 20 disordered photonic crystal sections of the superlattices – each of the supercells having the same level of randomness but a different disorder realization. Furthermore, in the numerical modeling results, we note that in the high-index physical setting studied here we described the optical beam propagation with the 3D vectorial Maxwell's equations instead of using a Schrödinger-type equation to account for the wave dynamics. The measured bandwidth increase of nearly-zero ν with increasing disorder is also supported by our 3D simulations, both in terms of the general wavelength dependence of ν and its estimated bandwidth from disorder. We note the localization bandwidth computation is a higher-order analysis, especially with the disorder lattice models of ~ 4,000 or more scattering sites where there are slight deviations between the exact numerical and experimental lattice instances and with the experimental samples containing additional disorder sources (such as from the sidewall roughness) that can account for the measured larger bandwidths.

This phenomenon of disorder-induced enhanced beam collimation is reminiscent of transverse localization. In isotropic media, ballistic transport is characterized with ν = 1 and diffusive transport is characterized with ν = 1/2; our measurements and simulations clearly demonstrate that the photon transport is arrested by disorder with ν values predominantly less than 0.05 in our disordered superlattices, even exceeding that of circular regular lattices. For the largest disorder (THS), we observed the strongest localization, with consistently near-zero ν values, averaged at 0.017, and with an almost flat spectral dependence of ν. In this regime for all disordered superlattices, the beam is localized and its divergence is arrested by the structural disorder in the superlattices, subjected only to statistical fluctuations in the scattering sites. The observed transverse localization arises from multiple coherent scattering of light induced by the disordered potential, forming localized guided resonances at the homogenous – photonic crystal interfaces. We also note that fluctuations in $\omega_{FWHM}$ increases with disorder (images in Supplementary Information V) and is consequently inversely proportional to the dimensionless conductance $\Sigma_o$ ($\Sigma_o \approx \rho(\omega) \cdot D_o \cdot l^*$, with the diffusion coefficient $D_o \approx$ the power exponent slope $\nu$). This reduced dimensionless conductance for increasing disorder arises due to coherent backscattering in the guided resonances over macroscopic lengthscales and, hence, reasonably the larger $\omega_{FWHM}$ fluctuations.

We further confirm the transverse localization through the analysis of the transverse intensity beam profile and its *z*-axis spatial progression. This is performed by examining the transverse intensity profile fitted to an exponentially-decaying form $I \sim \exp(-\frac{2|x|}{\xi})$ where $\xi$ is the localization length (the exponential decay length of the confined modes and defined with $l^* exp\,(\pi\, k_T\, l^*/2)$, a characteristic lengthscale of Anderson localization. For instance, in the case of HHS investigated at $\lambda_{ec} + 17$ nm, our analysis shows that $\xi$ is 2.5 μm for the $z = 25$ μm location and it roughly preserves an exponential transverse profile (instead of a Gaussian profile as is the case for diffusive transport). We also note that the beam profile becomes increasingly more asymmetric as disorder increases (as detailed in the Supplementary Information V, Figure S12). The exponential profile is also found to be the best fit for the SHS and THS cases. The exponential profile is a clear indication of chip-scale localization, with wave interference from the interplay of disorder on the periodic lattice.

For the circular superlattices without appreciable disorder, a new type of anisotropic medium based on cascaded excitation of guided resonances is observed with highly dispersive features and supported by both experimental measurements and numerical modeling. With increasing disorder, beam collimation in heptagonal, square, and triangular superlattices are observed for the first time. With increasing disorder strength, we observed increased collimation bandwidth, tighter collimation than in regular circular superlattices, and enhanced transverse localization. Transport in disordered superlattices reaches a regime of almost arrested diffraction, departing significantly from diffusive and ballistic transport, a phenomenon verified by the power-law scaling of the beam width and exponentially-decaying asymmetric intensity beam profiles in the localization regimes.

The observed transverse Anderson localization allows us to access values of the collimation bandwidth that is difficult to access through other approaches. With analogy to electronic transport, these observations allow us a means to probe the transverse Anderson localization of photons in solid-state semiconductors, including the role of guided resonances and the localization evolution. Future studies include optical nonlinearity perturbations to the localization (for different disorder levels) with potential spontaneous pattern formation, background scattering potentials with quasicrystal geometries, or the probing of these spatial localized modes with entangled biphoton states. The optical measurements developed here can find applications to other areas of physics as well. For example, photon transport in our superlattices is in many aspects analogous to electron wave dynamics in graphene heterostructures [26], so that similar effects could be observed in electron transport in a superlattice of closely-spaced disordered graphene nanoribbons [27]. The role of the guided resonances in this system arises from the nanoribbon edge states. These same phenomena could also be explored in other studied electronic systems such as semiconductor superlattices [28] and oxide heterojunctions [29], with interface states playing the role of guided resonances. Matter wave transport in atomic [30-31] and polariton [32] Bose-Einstein condensates trapped in suitably designed optical superlattices could also provide fertile testing ground of the conclusions of our work.

**Methods**

**Device nanofabrication:** The photonic crystal structures shown in Figure 1 were fabricated on a silicon-on-insulator wafer with a single-crystal silicon slab ($n_{si}$ = 3.48) with 320 nm thickness on top of a 2 μm thick layer of buried oxide ($n_{SiO_2}$ =1.46), with electron-beam lithography. ZEP520A (100%) resist was spin-coated at 4,000 revolutions per minute for 45 seconds to a thickness of ≈ 350 nm, and baked at 180 °C for 3 minutes. The JEOL JBX6300FS electron-beam lithography systems at Brookhaven National Laboratory-USA and ELIONIX ELS7500EX at National Cheng Kung University-Taiwan, respectively, were used to expose the ZEP520A resist to define the pattern, followed by development in amyl acetate for 90 seconds and rinsed with isopropyl alcohol (IPA) for 45 seconds to completely remove the residue of developer amyl acetate.

Oxford Instruments Plasmalab 100 was used for pattern transfer onto the silicon layer of the silicon-on-insulator wafer, using an inductively coupled plasma reactive ion etcher (ICP-RIE) to perform the cryogenic silicon etching. $O_2$ at -100 °C was applied in the chamber at first for cleaning and cooling, followed by cryogenic etching at -100 °C utilizing a mixture of $SF_6$ (40 sccm) and $O_2$ (15 sccm) at 15 W radiofrequency (r.f.) power, 800 W ICP power and 12 mTorr pressure for a total of 16 seconds. The resulting wafer was subsequently placed in a *n*-methyl pyrrolidone (NMP 1165) resist remover for about 4 hours to have the remaining of the ZEP resist completely removed.

**Band structure and time-domain numerical simulations:** The band diagrams of the photonic crystals and photonic superlattice are computed with RSoft's BandSOLVE, commercially available software that implements a plane wave expansion algorithm. In all numerical simulations a convergence tolerance of $10^{-8}$ was used to compute the frequency bands. The photonic bands of the photonic crystal have been divided into TM-like and TE-like polarizations, according to their parity symmetry. The effective refractive indices corresponding to the TM-like bands are determined from the relation $k= \omega |n|/c$, with ***k*** in the first Brillouin zone (see Supplementary Information).

The numerical simulations of the intensity field distribution have been performed by using MIT's MEEP, a freely available code based on the finite-difference time-domain (FDTD) method. In all numerical simulations we used a uniform computational grid with 33 grid points per micrometer. This ensures that the smallest characteristic length of the system (in our case, the hole diameter) is sampled by at least 10 grid points. In our FDTD simulations we used a cw excitation source of the same transverse size as the input waveguide, placed at the output facet of the waveguide.

**Acknowledgments**

We acknowledge discussions with Michael Weinstein, Serdar Kocaman, and Ching-Ting Chen. We also thank Shih-Ning Chiu on the data analysis, and the reviewers for helpful suggestions in the final manuscript version. This work is supported by the Office of Naval Research under Michael F. Shlesinger (N00014-14-1-0041) and the Studying Abroad Scholarship by the Department of Education in Taiwan. This work is also supported by NSF Division of Materials Research (1108176) and EPSRC EP/G030502/1. The electron-beam lithography carried out at the Brookhaven National Laboratory is supported by the U.S. Department of Energy, Office of Basic Energy Sciences, under Contract No. DE-AC02-98CH10886. The authors acknowledge the use of the UCL Legion High Performance Computing Facility (Legion@UCL) and associated support services in the completion of this work.

**Author contributions**

P.H., N.C.P., and C.W.W conceived the project. P.H. designed the photonic superlattices and performed numerical simulations, sample nanofabrication, measurements, and image analysis. C.C. and M.L. performed the electron-beam lithography, and C.C. performed the imaging with the focused ion-beam. J.F.M. performed the far-field and group velocity measurements. M.T. prepared the NSOM probes. N.C.P. designed the photonic superlattices and performed the theoretical analysis and FDTD numerical simulations. P.H., N.C.P., and C.W.W. wrote the manuscript, and all authors discussed the manuscript.

**Additional information**

The authors declare no competing financial interests. Supplementary information accompanies this paper online. Correspondence and requests for materials should be addressed to P.H. (email: ph2285@columbia.edu), N. C. P. (email: npanoiu@ee.ucl.ac.uk), and C.W.W. (email: cheewei.wong@ucla.edu).

**Figure captions**

**Figure 1 | Ordered and disordered superlattices. a,** Example of nanofabricated silicon photonic superlattices with 20 superperiods and single-mode input waveguide, imaged through focused ion beam. Inset: on-chip input waveguide with one-by-four splitter to four parallel superlattices for normalization. **b,** Ordered superlattices with circular holes. The dashed white lines depict the homogeneous region of the superlattices. **c,** Structural disorder is introduced by replacing the circular holes with heptagonal ones (≈ 2% structural disorder), and rotating them with an angle prescribed by a uniform random distribution. **d,** Square holes (≈ 6% structural disorder). **e,** Triangular holes (≈ 13% structural disorder). **f,** Band structure of the circular-hole superlattices, for the transverse wavevector component $k_x$. Inset: the flat bands (highlighted in red) near the normalized frequencies of ≈ 0.314 and 0.327 correspond to two guided resonances excited in the transverse photonic crystal waveguides (top and bottom insets respectively) with the computed $|E|^2$-profile plotted.

**Figure 2 | Dispersive-propagation numerical maps of the ordered and disordered superlattices. a-d,** The plotted effective beam width (blue with the tightest spatial extent; red the widest) is determined from the near-field spatial distribution of the field intensity, computed from 3D finite-difference time-domain numerical simulations. For the circular-hole superlattices (**a**), collimation is observed to be centered at 1550 nm. The heptagonal-hole (**b**), square-hole (**c**), and triangular-hole (**d**) superlattices show larger collimation bandwidths compared to the circular-hole superlattices. Input beam width is 450 nm.

**Figure 3 | Dispersive-propagation of the ordered and disordered superlattices. a-d,** The distribution of effective beam width versus wavelength at selected positions (color crossbars), for the circular-hole superlattices (**a**), heptagonal-hole (**b**), square-hole (**c**), and triangular-hole (**d**) superlattices, illustrating a flatter spectral response with increasing disorder.

**Figure 4 | High-resolution far-field infrared scattering images illustrating photon transport in the disordered superlattices. a,** Circular-hole, heptagonal-hole, square-hole, and triangular-hole superlattices (SL) at the $\lambda_{ec}$ wavelengths. **b,** Disorder media without superlattices and with

collimation mechanism solely from flat spatial dispersion surfaces. Measurements illustrated at the $\lambda_{ec}$ wavelengths in presence of disorder. Beam widths in these few hundred micrometer-scale media increase when increasing disorder, contrary to the superlattices.

**Figure 5 | Disorder-induced enhanced photon transport at the onset of transverse localization.** Schematic of infrared scattering for superlattices with different disorder (from left to right): circular, heptagonal, square, and triangular scatterers. The color plots correspond to the different wavelengths shown in the other panels. The beam diverges in the circular-hole superlattices but shows collimation in the heptagonal- and square-hole superlattices.

**Figure 6 | Photon transport enhanced by transverse localization. a-d,** Log-log plots of the experimentally derived effective beam width $\omega_{FWHM}(z)$ versus propagation distance for (**a**) circular-hole, (**b**) heptagonal-hole, (**c**) square-hole, and (**d**) triangular-hole superlattices, determined from the full-width half-maximum of the far-field infrared scattering, within the spectral region $\lambda_{ec}$ - 4 to $\lambda_{ec}$ + 24 nm. The red solid lines represent the approximate wavelength for most effective collimation $\lambda_{ec}$ (with least beam divergence; see also Supplementary Information V). Inset: distributions of slope ν within the spectral region $\lambda_{ec}$ - 10 to $\lambda_{ec}$ + 25 nm. The green filled dots and black open circles represent the experimental ν–datapoints and numerical ν–simulations respectively.

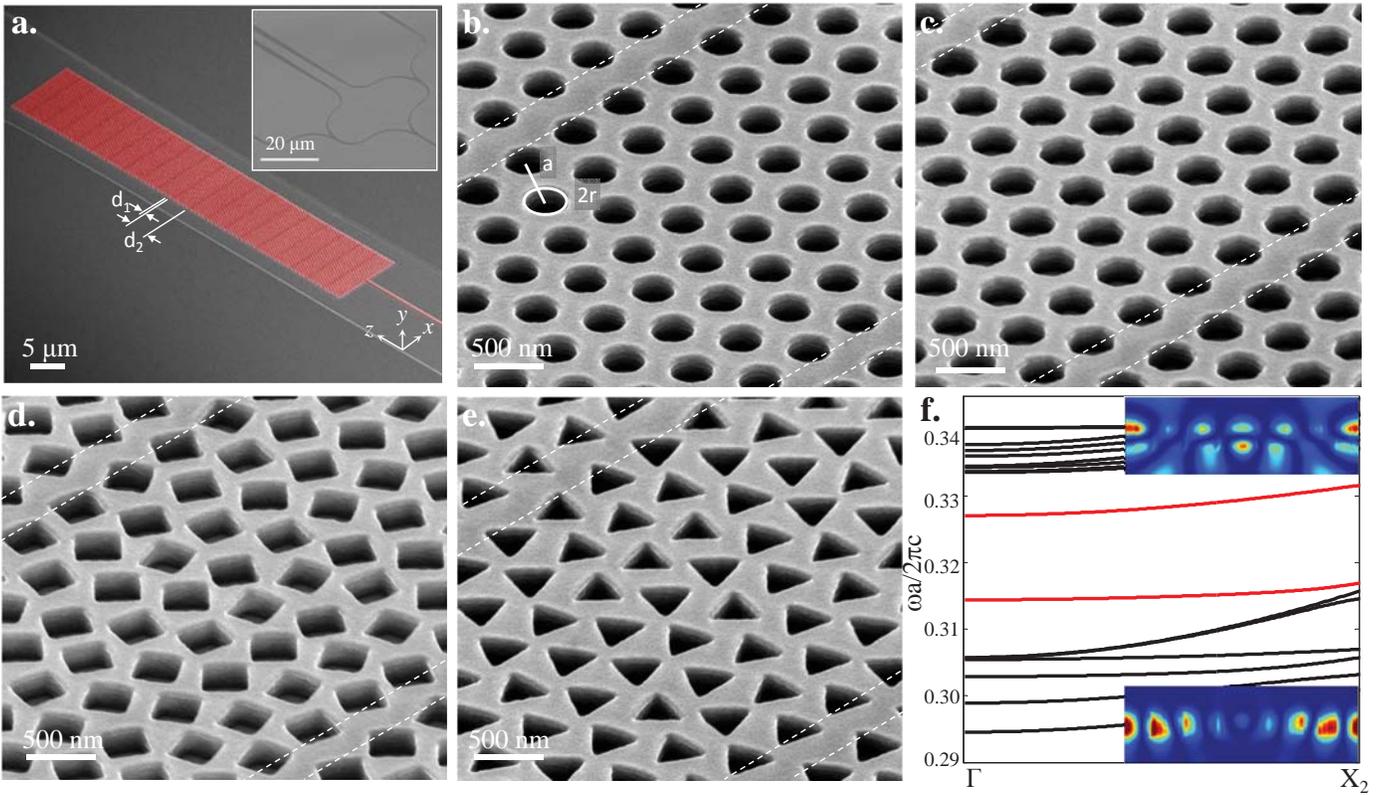

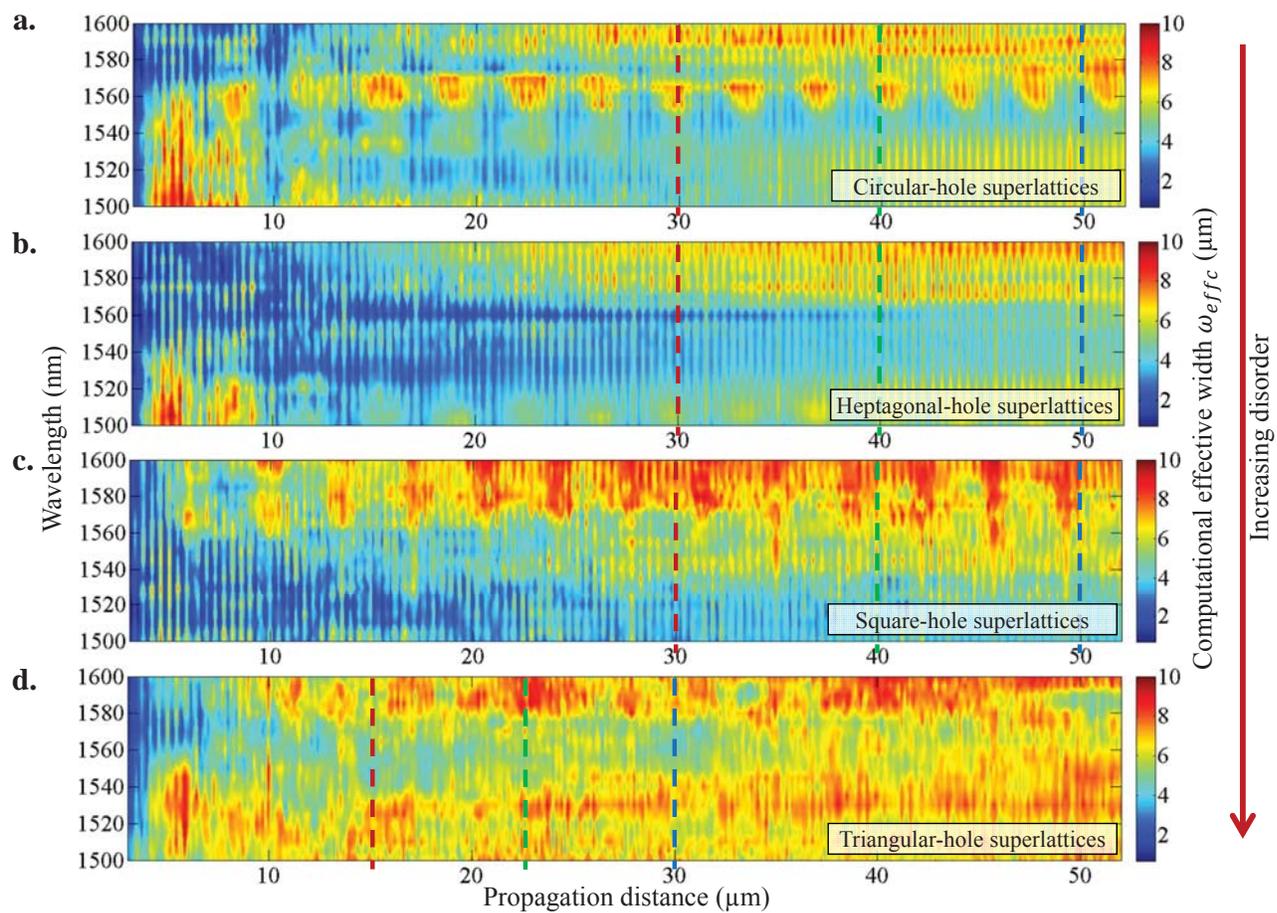

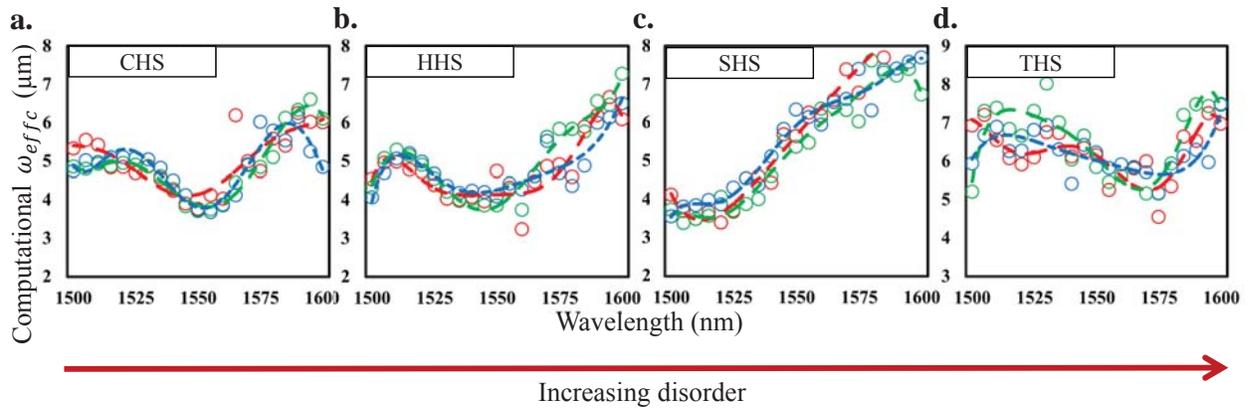

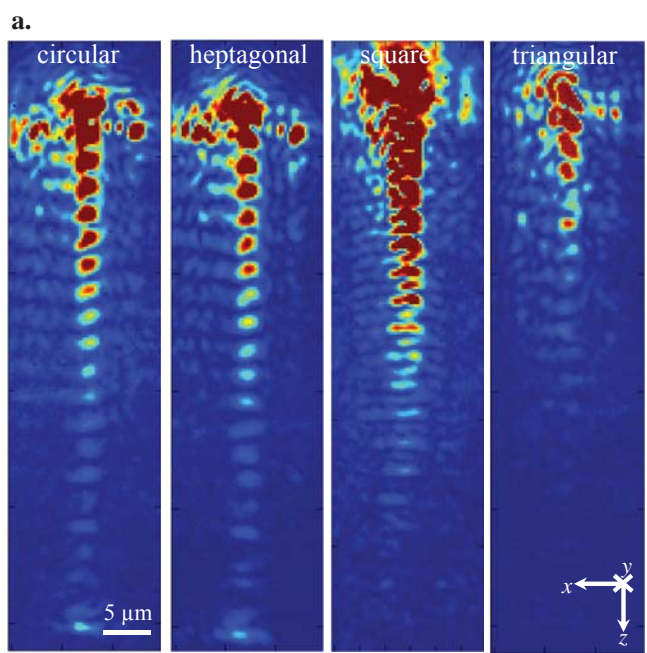 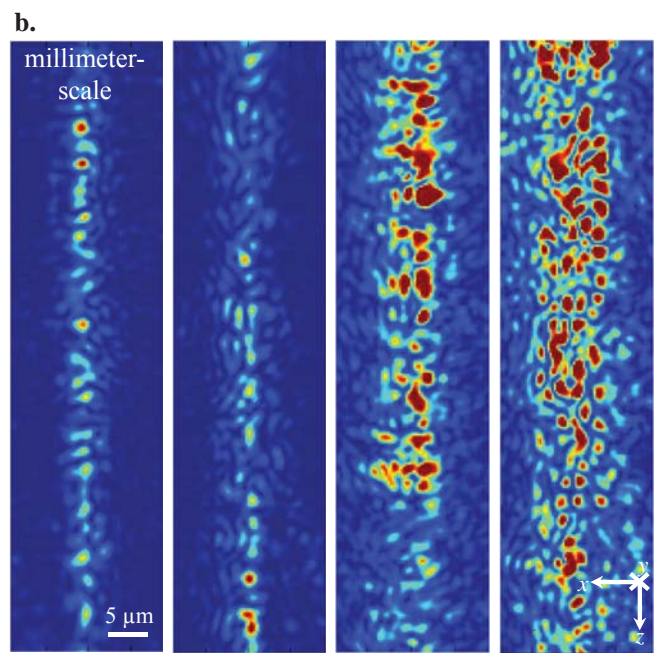

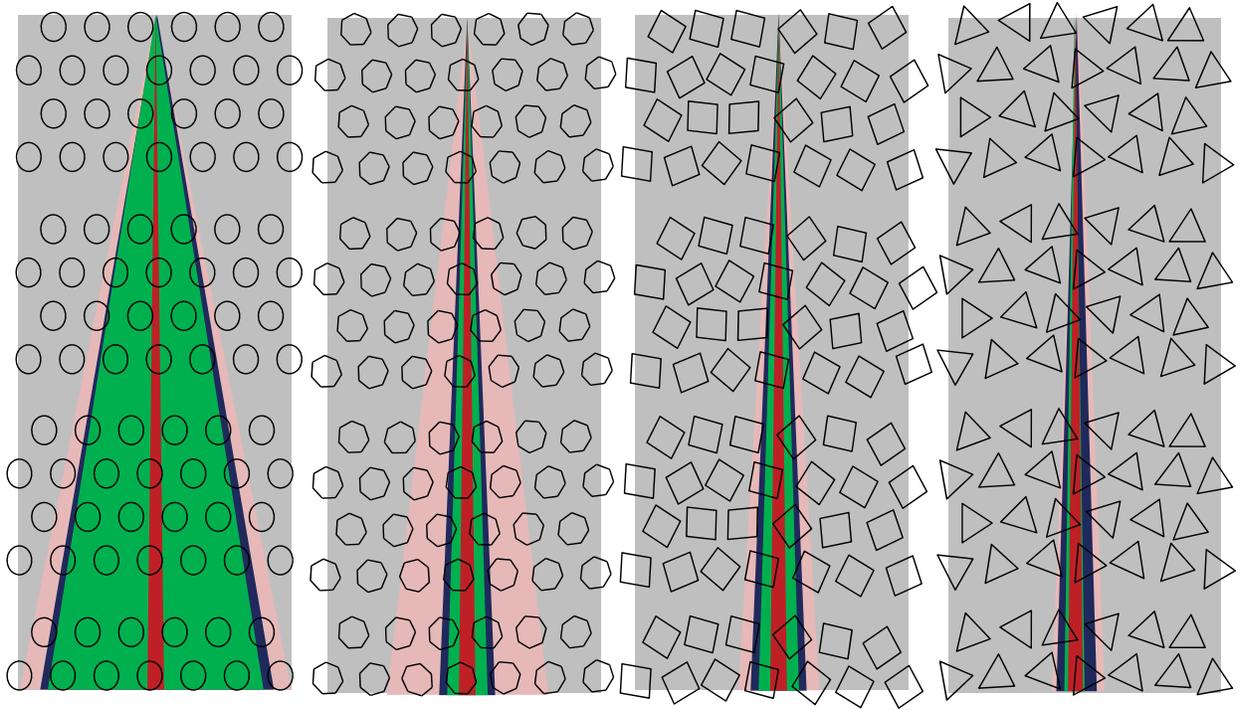

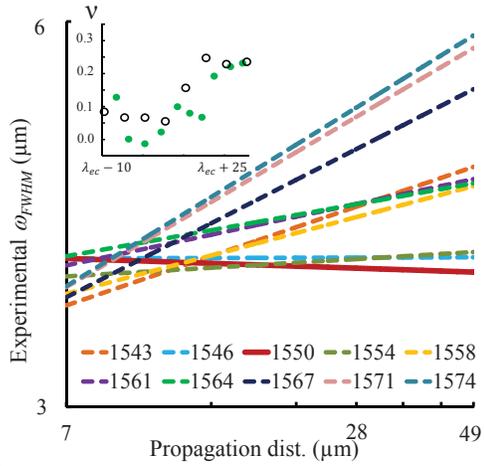

**b.** circular

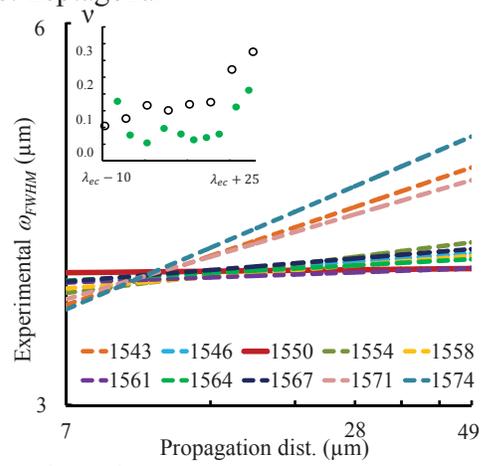

**c.** heptagonal

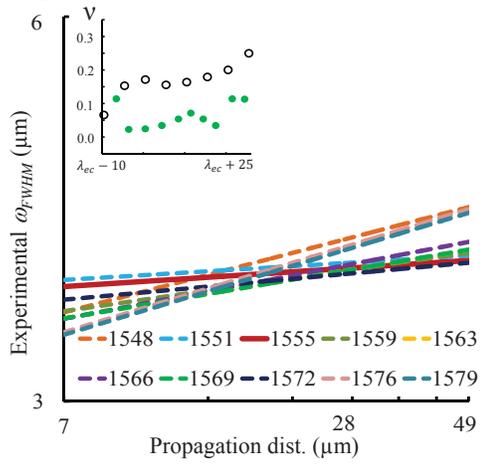

**d.** square

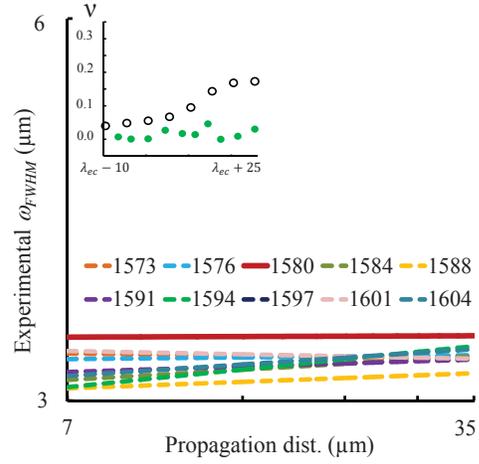

**e.** triangular